\documentstyle[12pt,aaspp4]{article}
\begin{document}
\title{Galaxy population properties in the rich clusters MS0839.8+2938, 
1224.7+2007 and 1231.3+1542\altaffilmark{1}}
 
\author{J.B. Hutchings, L. Edwards} 
\affil{Dominion Astrophysical Observatory\\Herzberg Institute of Astrophysics,
National Research Council of 
Canada\\ 5071 W. Saanich Rd., Victoria, B.C. V8X 4M6, Canada} 
 
\authoremail{john.hutchings@hia.nrc.ca}

\altaffiltext{1}{Based on observations with the Canada France Hawaii Telescope,
which is operated by the NRC of Canada, CNRS of France, and the University
of Hawaii}

\begin{abstract}
This paper discusses the galaxy populations of three rich clusters, with
redshift 0.19 (0839+29), 0.24 (1231+15), and 0.32 (1224+20), from the 
database of the CNOC1 consortium. The data consist of spectra of 52 
cluster members for 0839+29, 30 members for 1224+15, and  82 members for 1231+15, and there are comparable numbers of field galaxy spectra.
0839+29 is compact with no strong radial gradients, and possibly dusty.
1224+20 is isolated in redshift, has low velocity dispersion around the
cD galaxy, and low 4000\AA ~break. 1231+15 is asymmetrical and we discuss 
the possibility that it may be a recent merger of two old clusters. We 
find few galaxies in 0839+29 and 1231+15 with ongoing or recently truncated
star-formation.

\end{abstract}

\keywords{galaxies: clusters: individual (0839+29, 1224+20, 1231+15),
galaxies: evolution, galaxies: stellar content}

\section{Introduction and data}

This paper continues the discussions of galaxy populations of rich clusters
of intermediate redshift, based on the CNOC1 project database. The project
and basic data have been described by Yee, Ellingson and Carlberg (1996), 
and similar cluster
investigations have been published by Abraham et al (1996) and Morris et al 
(1998) for the clusters A2390 (z=0.23) and MS1621.5+2640 (z=0.42) (papers 1 
and 2). Further discussion of the whole cluster ensemble has been carried
out by Balogh et al (1999). We also discuss VLA radio imaging data for the
clusters 0839+29 and 1224+20, which were obtained in conjunction with the
CNOC1 program. The VLA data were obtained with the C configuration
at wavelengths 20cm and 6cm. Some details of the VLA observations are given
in paper 1. 

    The clusters selected for this paper were chosen from the CNOC database because of their
sample size, redshift comparison with those of papers 1 and 2, and the
radio imaging data. The investigation follows the same lines as papers
1 and 2: discussion of the spatial distribution of spectral properties
and colours, particularly with projected distance from the central galaxy,
and fitting of models that indicate the stellar content and star-formation
history of galaxies in the cluster. We also compare the properties of
different clusters.

   The data are described by Yee, Ellingson, and Carlberg (1996), and 
papers 1 and 2 contain full intoductory material on the galaxy population
studies: the reader is referred to those
publications for details. Our treatment of the data is the same as 
in papers 1 and 2. As in paper 2, we have used the GISSEL96 models for
comparison with the measurements from the data. Table 1 summarizes the
data and properties of the three clusters.

    The D4000 index is measured the same way as in the other cluster papers
(Hamilton 1995).
Balogh et al (1999) use a narrower definition in their measure which may be
less prone to systematic effects of redshift, passband, and reddening.
We have compared measures made with the `standard' passbands and two
reduced passbands of Balogh (private communication). For the clusters in
this paper, the mean ratio of narrow to standard passband measures
ranges from 1.00 to 1.06 for an intermediate bandpass, and from 0.85 to
0.90 for a very narrow one. We do not think this indicates any significant
systematic effect in retaining the original Hamilton passbands, and 
this has the advantages of enabling direct comparison with the two other 
CNOC cluster papers.  As in the other papers, the same measures
are made in the model spectra as the observed.

We did look for systematic differences with
Balogh et al's results by counting galaxies in the boxes they define in the
D4000/H$\delta$ plane. Balogh et al count galaxies in these boxes 
in order to compare their results with those of Barger
et al (1996). The boundaries chosen pass through the centre of the dense
distribution of galaxies, so that the numbers are very sensitive to the
exact placement of the boundaries. Using the Hamilton D4000 measure, we can
closely match the counts of Balogh et al with the D4000 boundary moved from
1.45 to $\sim$1.65, which is also close to the variation introduced by
bandpass choice, as noted above.
In view of the sensitivity of number counts to the 
chosen boundaries, and the relatively small numbers of galaxies in our 
clusters, we do not think such number counts are particularly useful for
this paper. With the above D4000 normalisation, they are however, the same 
as those of Balogh et al for
the whole CNOC cluster sample, within the poisson scatter expected for the number counts in the individual clusters in this paper.

  The H$\delta$ measures have also been the subject of discussion. Abraham
et al (1996) used a combination of two different measures intended to account
for the changing contamination of the continuum bands with different stellar
populations. Morris et al (1998) noted that since the same measures are made 
on model
and observed spectra, and the conclusions are unaffected, only one of the
H$\delta$ measures was used. However, a correction of 1.7A was applied to 
account for the different resolution of the model and observed spectra.
Most recently, Balogh et al have claimed that the sampling has little
effect on the H$\delta$ measure, and omitted this correction. Rather than add 
to this debate, in this paper, we have followed Morris et al to enable 
direct comparison with their results.

  Balogh et al (1999) have found that the formal uncertainties in some
line indices underestimate the reproducibility of their measurements.
For comparison with the cluster papers 1 and 2, we have not applied
any correction factors to the errors we display. The mean scale factors to
these error bars derived by Balogh are 2.0, 1.4, and 1.4 for D4000, 
H$\delta$, and [O II], respectively.

   For each cluster we discuss membership and define a sample of field
galaxies from the same sample. There are a few galaxies which lie at the
edge of the clusters which may not (yet) be bound members, and these are
designated near-field galaxies, as in paper 2. We use the same definition
of these as Balogh et al, although they stand out clearly to the eye
in the redshift-radius plots. Galaxy colours from the direct
images and spectral indices (D4000, H$\delta$, and [O II]) from the spectra
are plotted
with clustocentric distance. The galaxies are grouped in radius bins and
the measured quantities of each subset compared with the same measures on
stellar population models for galaxies. We present mean properties of
various subsets of the galaxies, and present mean spectra for several of these
subsets. We also show the spatial distribution of some subsets of the galaxy
populations.

   We discuss the inferred stellar populations and galaxy evolution for
the clusters and compare the results for all the CNOC clusters thus analysed.

\section{0839+29}

   This cluster (along with 1224+20) has been discussed as a cooling
flow cluster on the basis of extended H$\alpha$ emission, by Donahue,
Stocke, and Gioia (1992), and from X-ray imaging by Nesci, Perola,
and Wolter (1995).

   The membership of the cluster from the sample spectra is illustrated
in figure~\ref{z0839}. Figure~\ref{rad0839} shows the colour gradient
with magnitude and radius, and from this we adopt the red sequence of
members at g-R$>$0.75. The blue and red galaxies fall within the
same cusp envelope in Figure~\ref{z0839} (which seem to be poorly sampled at
radii above $\sim$400"), so the colour cut is not particularly significant.
We have labelled 3 galaxies that lie outside a symmetrical cusp envelope
but very close to the cluster, as `near-field'. These are all blue, as was
found for similar galaxies in 1621+26.

   The cluster is relatively compact within the large size of fields sampled
(see Table 1). Only 
two galaxies with the cluster redshift are seen further than 463" from the
central galaxy, compared with 1400" for the similar redshift A2390,
and 700" for the higher redshift 1621+264 in papers 1 and 2. Carlberg,
Yee, and Ellingson (1997) compare the dynamical parameters of the CNOC
clusters, and 0839+29 has the lowest virial radius of all, except 
for the binary cluster 0906+11. 

   The spatial distributions of some subsets of the galaxies are shown
in Figure~\ref{spat08}. The field galaxies used for comparison lie in the
redshift range 0.17 to 0.22. There are 25 of these and their redshift range 
is small enough that no colour corrections are needed for model comparisions.
In any case, in this paper we do not show any color models.

   As in papers 1 and 2, we choose the colour, D4000, H$\delta$, and [O~II]
spectral measures for discussion and modelling for stellar populations.
Figure~\ref{rad08} shows these plotted with (log) radius, and compared with the non-member sample. The measurements and error estimates were made using the 
same algorithm as in papers 1 and 2. Table 2 shows mean properties of various
subsets of the galaxies. This table also includes indices of other lines
that are sensitive to stellar population. The table also includes mean
values for extreme [O II] and H$\delta$ as defined. Such quantities are
shown for all three clusters as different ways of summarizing the data.

   There is a group of 15 galaxies with redshift near 0.217 that appear to
be clustered. They are located close to each other in the sky, centred 
about 100" from the centre of 0839+29. Table 2 also shows their properties:
they seem to have active star-formation and a young stellar population. 
There is a group of 4 galaxies that lie at an intermediate redshift between 
the main and this small cluster, that has even younger populations. These
appear as a group in the sky, but some 200" to the other side of the
main cluster centre. Thus, they are not obviously involved with either
cluster, but may be a small group on their own.

  In Figure~\ref{rad08} we have plotted measurements with log radius: the cD
galaxy  is arbitrarily placed at 
0.41, since it lies at zero radius. The plot shows an apparent gradient of 
D4000 with radius in the cluster, particularly beyond 250". 
In all the data the [O~II] emission line is not seen below redshift 
$\sim$0.2 because of the spectral bandpass, but other field samples
and those above z$\sim$0.2 show no trend across the cluster redshifts that
might concern our camparison with cluster members.
The suggested smaller cluster at z=0.217 does stand out in [O~II] behaviour.
The colour plot shows little change in the red sequence,
and there are a few H$\delta$ strong galaxies seen in the central part of
the cluster. The majority of the cluster population seems to have evolved
passively for some time, and there is little accretion happening. The three
near-field galaxies do stand out as younger and forming stars. One of them
has an Seyfert spectrum, whose H$\delta$ emission is off-scale in the plot.
We note that the spectral passband prevents [O~II] measures in the lower
redshift field galaxies.

   Figure~\ref{spec08} shows mean spectra of the major subgroups. These are
derived from the galaxies with best signal in each group, but we have
checked that their mean measured properties are still representative of
the whole subgroup. These may be compared with similar mean spectra from
papers 1 and 2, and the other clusters in this paper, but have no remarkable
properties in that context.

   We have divided the members into three radial distance groups of roughly
equal population, and with boundaries close to radii where measured quantities
change. Figure~\ref{mod08} shows plots of measured quantities
for these subsets, and also the field sample. The diagrams also show
models computed from GISSEL96, as in paper 2.    The upper curve is for
a 1 Gyr starburst followed by passive evolution, and the lower curve is
for exponentially decreasing star-formation with a timescale of 4 Gyr.
Both models are for solar abundance, and the H$\delta$ index is corrected
for spectral resolution as in paper 2. Paper 2 also discusses in more
detail the effects of different abundances and initial mass funtion.

These plots show there are several blue galaxies with fairly strong
H$\delta$ in the central region (or projected
on to it, as their high velocity dispersion suggests). This is unusual in
the central parts of these rich clusters. There are also a few galaxies with
ongoing star-formation in the outer regions. As noted from the other
plots above, there is an old population seen at all radii among the members.

   In Figure~\ref{simd} and Figure~\ref{simhd} we show the variation of
D4000 and H$\delta$ with age together with the histograms of these
quantities in all three clusters. At the redshift of 0839+29 we expect
the oldest populations to be some 12 Gyr. The bulk of the cluster population
seems to be signficantly younger than this, and from figure~\ref{mod08}
we see little radial gradient.
The outer cluster has no old populations. The field galaxy population
is similar to the outer cluster members. The inner cluster has several galaxies
with higher H$\delta$ index than the field or the models, as seen in 1621+26.
This may result from a truncated IMF in the inner cluster galaxy populations,
as noted in paper 2.
   
   We find 4 galaxy coincidences with radio sources in the VLA images.
These are summarized in Table 3. The cD galaxy is the strongest source.
The second radio source is the Seyfert galaxy. It is a disk galaxy with a
bright nucleus. The next source is a red galaxy within the central cluster,
with a fairly young stellar population. It is asymmetrical and possibly
interacting. The fourth galaxy has no spectrum but is clearly interacting
and is probably a foreground galaxy, judging by its size and brightness.
Table 3 summarizes the radio source identifications.

   In order to investigate the galaxy environments of cluster members,
we created a table of the distance and magnitude of galaxies within 100 pixels
($\sim$33 arcsec) of each, from the imaging catalogue of the field,
which extends to magnitude 24 and fainter. Plots
from these numbers do not reveal any connection between spectra and apparent
small groups. The most clustered members do not have unusual spectral
measures or active star-formation.

\section{1224+20}

   The data for this cluster are published by Abraham et al (1998).
This cluster has been discussed as an example of velocity bias by
Carlberg (1994), and Fahlman et al (1994) also find a weak lensing mass
that exceeds the virial mass by a factor of more than 2. 
Lewis et al (1999) report an X-ray mass estimate which is consistent
with the dynamical one. Carlberg, Yee
and Ellingson (1997) show from the CNOC database that the cluster radius
is lower than average for the group, and the velocity dispersion close
to average at 798 km s$^{-1}$. In addition to the CNOC database, we
also have VLA maps at 20cm and 6cm of the cluster.

    The CNOC spectroscopic database is relatively small for this
cluster, with 75 spectra, of which 30 are considered members. Among the
sample, there is a very clear redshift separation of members and field
galaxies, but the sample does not indicate the usual central velocity
spread among the members. The colour plots in Figure~\ref{col122}
suggest the red sequence colour cut-off at g-R=0.95: however, with this
cut, the blue member galaxies do not show the usual larger scatter
in redshift-radius distribution than the red members
(see Figure~\ref{z122}). It may be the lack of central velocity spread 
among the CNOC members that causes the virial mass discrepancies noted 
above. The densest grouping of cluster members occurs some 80 arcsec
away from the cD galaxy nominally regarded as the cluster centre. 
This dense group
does not have a special redshift, and the low velocity outliers do
not form a spatial group, so there is no obvious simple substructure.
Figure~\ref{spat122} shows the distribution of galaxy subsets in the sky.

   The radial gradients and field comparisons of measured quantities are
shown in Figure~\ref{lograd122}. The cluster members generally have 
considerably more evolved stellar populations than the field. Some of
the blue members do have younger populations, but do not show much
star-formation. The values of D4000 however, are unusually low for
an evolved population, with an average for red members of 1.7, compared
with values well over 2 for other CNOC clusters (see figure~\ref{simd}). 
The measures were made with the same code as all other spectra, and are 
small with any definition of D4000 index and redshift, so this suggests that 
the cluster may have formed fairly recently, and that star-formation turned
off not very long ago.

   The mean spectra of representative subsets are shown in 
Figure~\ref{spec122}. The low D4000 break is also seen here. Table 4 
gives the mean measured values. We looked at the off-centre compact
group (`A') separately, but they do not show any interesting difference
that suggests they are a subgroup, as also noted above.

   Figure~\ref{mod122}
shows model measures and the data in two radial bins, and the field. There
is little radial gradient in D4000 which is also consistent with recent
cluster-wide evolution in this quantity shortly after star-formation ceases. 
The field
galaxies in this region have very blue colours, presumably indicating young
stellar populations. The member
and field galaxy colours are consistent with this suggestion. There is
little or no colour gradient in the red sequence within the cluster. 

  In view of the virial peculiarities noted for the cluster, the low
velocity dispersion about the cD galaxy, and the dense grouping of
members off the centre, we investigated the consequences of adopting
this dense region as cluster centre. The redshift/radius relationship 
does look more normal: it is highest at this centre, although still low.
This centre also shifts the two low D4000 (young population) galaxies
out of the central 100" radius bin, which is also more normal. The second
brightest member galaxy in our sample (18.6 compared with 18.0 for the cD)
lies near this new centre, at +91, -47 in arcsec from the cD. However, its
redshift is far from the cluster mean at 0.3198, and it has an old population
and red colour. The dymanic situation of this cluster needs more detailed
investigation. 

   Among the radio sources in the field, only four correspond with galaxies
in the image. The brightest two sources (15 - 20 mJy at 20 cm) are large
foreground galaxies for which we have no spectrum, or hence redshift. The 
next source has flux 2.6 mJy at 20 cm, and is a 16mag foreground galaxy
at redshift 0.05. The last source identified has 20cm flux 0.4 mJy, and 
is measured at 3.8" from the cD galaxy, which we adopt as the identification.
Thus, there is only one weak cluster source which is the cD. 

   Searching the photometric catalogue for faint companions, we find
the crowded region at radius 65" - 85" also stands out down to 24th magnitude.
In addition, there are few faint galaxies inside this radius. This suggests
that the region around the cD galaxy (if it is the cluster central region) 
has been cleared of 
faint galaxies or has enough extinction to hide background galaxies.

\section{1231+15}

   This cluster has a larger radius and lower velocity dispersion than the
other two. The sample includes three masks in the central field, and one
each to the north and south, so that the central region is well sampled.
The red sequence cutoff is easily defined 
at g-R=0.7, as there is a wide gap with no galaxies of colours between 0.62 
and 0.76 (Figure~\ref{color123}). The red sequence colour is normal for its
redshift among CNOC clusters, and has a clear gradient with radius and 
magnitude. 
The redshift-radius plot (Figure~\ref{z123}) shows a well 
virialised distribution of galaxies, both red and blue, with 3 near-field
galaxies (in the radius range 250-300 arcsec) which are red. 
These do not form a close physical group, in spite of the
very similar redshifts.  
   The cD galaxy has a velocity some 250 km s$^{-1}$ lower than the
cluster mean. The galaxy distribution on the sky is asymmetrical about
the cD (Figure~\ref{spat123}), suggesting that possibly the cluster has 
merged from two. We investigated this by splitting the galaxies into
groups, as shown in the diagram. The redshift-radius plots are very 
similar for the two, but group 1 has younger population indicators.  
Table 4 shows differences in this sense in all mean indices, each at 
about 1$\sigma$ level.   

   The plots of measured properties with (log) radius are in 
Figure~\ref{lograd123}. The cluster is characterised by high values
of D4000 (along with other line measures indicating old or high
abundance populations). This indicator is higher for group 1 than group 2.
The radial gradient of spectral features seen in Figure~\ref{spat123}
is unusual in showing undulations, and this too may be explained by
a different population (and hence history) of galaxies in the two main
groups. Figure~\ref{spec123} shows mean spectra of representative subsets.
   
    Figure~\ref{mod123}, \ref{simd}, and \ref{simhd} show the comparison
with models. This cluster has a significant number of high H$\delta$
red galaxies, and high D4000 values, particularly in the mid-range of
radius. This could arise from a higher abundance in the stellar populations,
possibly by being formed from galaxies that had long periods of star
formation before forming the cluster. There are similar galaxies in the field,
as well as some (still the majority) that have ongoing star-formation. 

   Looking for faint companions to cluster galaxies, we find that many of
the 82 have several galaxies within 5 arcsec, and 13 of them look like
compact groups in the sky. Most of those are unremarkable red galaxies
but there are three with blue colours and/or active star-formation. Of
those, the bluest is in a populous group of 10, while the other two
are have 3-4 very close companions. In these cases at least it seems
possible that the stellar populations are affected by local interactions
within the groups. This cluster has a higher density of faint galaxies
than the other two. The average number of galaxies to the catalogue limits
are 45 for 1231+15, 20 for 1224+20, and 15 for 0839+29. Comparing the
magnitude distributions, the 0839+29 catalogue is less complete to 24 mag, 
while the 1231+15 is the most complete. The distributions of slit masks
across the three clusters are similar (Table 1), but less centrally
concentrated in 0839+29. However, the groups of around the star-forming
member galaxies in 1231+15 are not at the faint limit and would be visible
in either of the other two catalogues.

\section{Comparison and discussion}

   We have presented the same plots and model comparisions for the three
clusters. We have not done this in the same detail as in papers 1 and 2,
focussing principally on H$\delta$ and D4000 in the model discussions.
Other quantities are shown in the data plots and tables. 

   The colours of the red sequences for the clusters and those of papers
1 and 2 (A2390 and 1621+26) follow a progression with redshift, except
for 0839+29 which is redder by some 0.15 in g-R. This may indicate the
presence of some dust in this cluster, since the line indices do not
correspond with high abundance or age. This cluster may contain populations
of different ages or abundance, as the D4000 distribution appears bimodal
(see figure~\ref{simd}).
This is seen in 1621+26, but in 1621 the main population is old/high abundance
while in 0839+29 it is young/low abundance.

   The cluster 1224+20 has remarkably low D4000, suggesting that its stellar population is either of low abundance or that the cluster itself has
only formed, and hence stopped star-formation, recently. As noted, the
discussions of velocity suggest that the cluster may be in an early stage of
its dynamical evolution too. The highest concentration of member galaxies
and the largest velocity spread is found some distance from the cD galaxy.
The lack of star-forming members and strong
Balmer absorbers also suggest that there is little current infall, since
infall is regarded as truncating or restarting star-formation. The lack
of neighbouring redshifts among the field sample also is consistent with
this scenario.

  The cluster 1231+15 has a wide spread of age/abundance, and a complex
gradient of D4000 with radius. As we have discussed, there are suggestions
of a recent merging of major subclusters to form this cluster. The relatively
high fraction of star-forming and H$\delta$ strong galaxies in this cluster 
may also result from this.

   We are grateful to David Schade and Simon Morris for help with the
CNOC database and with the GISSEL models. 

\newpage

\centerline{\bf References}

Abraham R.G. et al 1996, ApJ, 471, 694 (paper 1)

Abraham R.G., Yee H.K.C., Ellingson E., Carlberg R.G., Gravel P., 1998,
ApJS, 116, 231

Balogh M.L., Morris S.L., Yee H.K.C., Carlberg R.G., Ellingson E., ApJ 
(in press: astro-ph 9906470)

Barger A.J. et al, 1998, ApJ, 501, 5223.

Carlberg R.G., Yee H.K.C., and Ellingson E., 1997, ApJ, 478, 462

Carlberg R.G., 1994, ApJ, 434, L51

Donahue M., Stocke J.T., Gioia I.M., 1992, ApJ, 385, 49

Fahlman G., Kaiser N., Squires G., and Woods D., 1994, ApJ, 437, 56

Hamilton D., 1985, ApJ, 297, 371

Lewis A.D., Ellingson E., Morris S.L., Carlberg R.G., 1999, ApJ in press
(astro-ph 9901062)

Morris S.L., et al, 1998, ApJ, 507, 84 (paper 2)

Nesci R., Perola G.C., Wolter A., 1995, A\&A, 299, 34

Yee H.K.C., Ellingson E., and Carlberg R.G., 1996, ApJS, 102, 289 

\newpage

\centerline{\bf Captions to Figures} 
   
\figcaption[hutchings.fig1.ps]{Members of cluster 0839+29 in the sample.
The nearfield galaxies may not be virialised members but lie far from
the nearest field galaxies. \label{z0839}}

\figcaption[hutchings.fig2.ps]{Plots of colour with magnitude and clustocentric
radius. We adopt a colour cut of 0.75 to define the red sequence of members.
\label{rad0839}}

\figcaption[hutchings.fig3.ps]{Distribution of 0839+39 field galaxies in the
sky. \label{spat08}}

\figcaption[hutchings.fig4.ps]{Properties of 0839+29 member and field
galaxies with projected distance from cluster centre. \label{rad08}}

\figcaption[hutchings.fig5.ps]{Representative mean spectra of subgroups
in 0839+29 field. Spectra are derived from 6-10 galaxies with good signal
and properties close to the subgroup means. They are shifted to the mean
cluster redshift and smoothed. \label{spec08}}

\figcaption[mod08.eps]{H$\delta$ and D4000 indices for 0839+29 members and
field galaxies, compared with the same measures from GISSEL96 models. The
dashed line shows passive evolution after a 1 Gyr starburst, and the solid line
represents exponentially decreasing star-formation with the timescale of
4 Gyr. \label{mod08}}

\figcaption[hutchings.fig5.ps]{Plots of colour with magnitude and clustocentric
radius for MS1224.7+2007. We adopt a colour cut of 0.95 to define the red
sequence of members. \label{col122}}

\figcaption[hutchings.fig6.ps]{Members and nearby galaxies in redshift space
around the centre of 1224+20. \label{z122}}

\figcaption[hutchings.fig7.ps]{Distribution of 1224+20 field galaxies in the
sky. \label{spat122}}

\figcaption[hutchings.fig8.ps]{Properties of 1224+20 member and field
galaxies with projected distance from cluster centre. \label{lograd122}}

\figcaption[hutchings.fig9.ps]{As Figure~\ref{spec08}, for cluster 1224+20.
\label{spec122}}

\figcaption[mod122.eps]{As figure~\ref{mod08}, for cluster 1224+20. The solid
lines are solar abundance models and the dashed line is for 1/5 solar.
\label{mod122}}

\figcaption[hutchings.fig10.ps]{Plots of colour with magnitude and clustocentric
radius for MS1231+1542. We adopt a colour cut of 0.70 to define the red 
sequence of members. \label{color123}}

\figcaption[hutchings.fig11.ps]{Members and nearby galaxies in redshift space
around the centre of 1231+15. \label{z123}}

\figcaption[hutchings.fig12.ps]{Distribution of 1231+15 field galaxies in the
sky. \label{spat123}}

\figcaption[hutchings.fig13.ps]{Properties of 1231+15 member and field
galaxies with projected distance from cluster centre. \label{lograd123}}

\figcaption[hutchings.fig14.ps]{As Figure~\ref{spec08}, for cluster 1231+15.
\label{spec123}}

\figcaption[mod123.eps]{As Figure~\ref{mod08}, for cluster 1231+15. Note
the large population of cluster galaxies with high H$\delta$ and D4000.
\label{mod123}}

\figcaption[simd.eps]{Growth of D4000 index from GISSEL96 models. The 
post-starburst models have passive evolution after a 1 Gyr starburst, and the
star-forming models decrease with 4 Gyr timescale.  The star-forming models
have very little change with abundance. The panels superpose the distribution
of D4000 indices for the members of the three clusters studied. \label{simd}}

\figcaption[simhd]{As Figure~\ref{simd}, but for H$\delta$. \label{simhd}}

\newpage
\footnotesize

\begin{deluxetable}{llllll}
\tablenum{1}
\tablecaption{CNOC data on clusters analysed}
\tablehead{\colhead{Cluster} &\colhead{z} &\colhead{Members} 
&\colhead{Field}  &\colhead{\# Fields\tablenotemark{a}} &\colhead{Comment}}
\startdata
0839+29 &0.19 &52 &99  &E,C(2),W1(2),W2(2) &VLA, Cooling flow\nl
1224+20 &0.32 &30 &45   &C(2) &VLA, Cooling flow\nl
1231+15 &0.24 &82 &78  &C(3),N,S &\nl

\enddata
\tablenotetext{a}{location of $\sim$10' fields in cluster, (with number of slit
masks)}

\end{deluxetable}

\begin{deluxetable}{lrlllrrcrrrr}
\tablenum{2}
\tablecaption{Mean properties of galaxies in 0839+29}
\tablehead{\colhead{Group\tablenotemark{a}} &\colhead{\#}  &\colhead{g-R}  &\colhead{D4000}
&\colhead{H$\delta$}  &\colhead{[OII]}  &\colhead{R\tablenotemark{b}}
 &\colhead{z}
&\colhead{m$_R$}  &\colhead{H$\gamma$}  &\colhead{CaK} &\colhead{g band}}
\startdata
Red  &38 &0.89 &2.07 &2.0 &-0.6 &132 &0.1935$\pm$0.0031 &19.5 &1.2 &11.7 &6.4\nl
Blue &12 &0.42 &1.51 &5.1 &-24 &254 &0.1933$\pm$0.0049 &19.9 &3.6 &2.3 &3.0\nl
H$\delta>$4\AA &14 &0.70 &1.89 &7.3 &-6 &134 &0.1942$\pm$0.0043 &19.8 &3.5 &9.1 
&5.3\nl      
[OII]$<$-15\AA &8 &0.46 &1.66 &4.8 &-38 &326 &0.1926$\pm$0.0048 &20.1 &2.3 &4.3 
&3.2\nl                                                    
Radius$<$85" &18 &0.82 &1.99 &2.2 &-6 &48 &0.1932$\pm$0.0045 &19.3 &2.5 &10.3
&5.6\nl
Radius 85-200" &20 &0.86 &2.04 &3.4 &2 &134 &0.1939$\pm$0.0028 &19.7 &1.1 &10.7
&6.2\nl
Radius$>$200" &12 &0.59 &1.69 &2.4 &-20 &378 &0.1944$\pm$0.0034 &20.0 &1.8 &6.0
&4.6\nl                                                             
Field  &25 &0.61 &1.66 &3.9 &-23 &(313) &0.17--0.22 &19.8 &1.0 &5.2 &3.3\nl
Field  &99 &0.72 &1.66 &4.7 &-27 &(385) &0.05--0.48 &20.6 &1.5 &4.6 &3.7\nl
Cluster\#2 &15 &0.63 &1.72 &4.0 &-16 &(263) &0.2167$\pm$0.0018 &19.7 &1.2 &5.8
&3.6\nl
\enddata                                        

\tablenotetext{a}{cD and Seyfert galaxies omitted;                                                                                                                     Blue/Red cut at g-R=0.75\\
~~~~~Line measures are in \AA}
\tablenotetext{b}{Mean projected radius from cD in arcsec}
\end{deluxetable}

\begin{deluxetable}{llclllll}
\tablenum{3}
\tablecaption{Radio sources in 0839+29 field}
\tablehead{\colhead{\#} &\colhead{m$_R$} &\colhead{" from cD}
&\colhead{RA(2000)} &\colhead{Dec(2000)} &\colhead{Opt-radio} 
&\colhead{S(20cm,6cm)} &\colhead{Notes} \\
&&\colhead{(RA, Dec)} &&&\colhead{(arcsec)} &\colhead{(mJy)}}
\startdata
1 &17.0 &0,0 &08:42:55.9 &29:27:27 &0.4 &21.2, 1.9 &cD galaxy\nl
2 &19.1 &-52.3, 58.3 &08:42:51.8 &29:28:25 &1.4 &3.7, 0.4 &Near-field Seyfert\nl
3 &19.5 &-26.9, 151.9 &08:42:53.9 &29:29:57 &1.1 &1.0, - &Member,
interacting?\nl
4 &19.5 &118, 42 &08:43:04.9 &29:28:08 &3.4 &0.8, - &Interacting, foreground?\nl
\enddata
\end{deluxetable}

\begin{deluxetable}{lrcllrrclrll}
\tablenum{4}
\tablecaption{Mean properties of galaxies in 1224+20 field}
\tablehead{\colhead{Group\tablenotemark{a}} &\colhead{\#}  &\colhead{g-R}  &\colhead{D4000}
&\colhead{H$\delta$}  &\colhead{[OII]}  &\colhead{R\tablenotemark{b}}
 &\colhead{z}
&\colhead{m$_R$}  &\colhead{H$\gamma$}  &\colhead{CaK} &\colhead{g band}}
\startdata
Red &23 &1.24 &1.69 &1.1 &0.3 &142 &0.3256$\pm$0.0037 &20.5 &2.0 &9.7 &6.4\nl
Blue &6 &0.63 &1.33 &4.3 &-13 &139 &0.3251$\pm$0.0028 &21.0 &0.9 &5.5 &3.4\nl
H$\delta>$3 &10 &0.87 &1.44 &5.3 &-8 &150 &0.3255$\pm$0.0032 &20.9 &1.1 &5.1 &4.0\nl  
[OII]$<$-5 &6 &0.90 &1.46 &4.4 &-15 &108 &0.3268$\pm$0.0023 &21.1 &0.0 &3.7 &3.1\nl 
Group A &6 &1.25 &1.78 &2.0 &1 &89 &0.3238$\pm$0.0041 &20.5 &1.9 &8.4 &7.2\nl 
Radius$<$100" &11 &1.14 &1.65 &2.2 &-2 &69 &0.3249$\pm$0.0034 &20.6 &3.0 &8.9 &6.7\nl
Radius$>$100" &18 &1.10 &1.60 &1.6 &-3 &186 &0.3258$\pm$0.0035 &20.5 &1.0 &8.8 &5.2\nl                               
Near field &6 &0.95 &1.41 &3.3 &-25 &240 &0.31--0.36 &20.4 &-0.6 &4.1 &3.9\nl 
Field &45 &0.82 &1.32 &4.1 &-27 &-- &0.05--0.61 &20.6 &-0.5 &7.5 &5.5\nl                                                  
\enddata
\tablenotetext{a}{Without central cD galaxy; Blue/Red cut at 0.95\\
~~~~~Line measures are in \AA}
\tablenotetext{b}{Mean projected radius from cD in arcsec}
                                                                                                                   \end{deluxetable}

\begin{deluxetable}{lrlllrrclrrl}
\tablenum{4}
\tablecaption{Mean properties of galaxies in 1231+15 field}
\tablehead{\colhead{Group\tablenotemark{a}} &\colhead{\#}  &\colhead{g-R}  &\colhead{D4000}
&\colhead{H$\delta$}  &\colhead{[OII]}  &\colhead{R\tablenotemark{b}}
 &\colhead{z}
&\colhead{m$_R$}  &\colhead{H$\gamma$}  &\colhead{CaK} &\colhead{g band}}
\startdata
Red &67 &0.89 &2.28 &2.2 &-1 &201 &0.2351$\pm$0.0025 &19.8 &0.6 &11.1 &6.7\nl
Blue &15 &0.40 &1.51 &6.4 &-26 &296 &0.2347$\pm$0.0026 &20.2 &0.4 &5.9 &2.7\nl
H$\delta>$5 &22 &0.68 &1.92 &7.9 &-8 &197 &0.2347$\pm$0.0027 &20.3 &-0.5 &10.0 &4.5\nl
[OII]$<$-20 &13 &0.49 &1.55 &5.0 &-38 &331 &0.2356$\pm$0.0025 &20.6 &-0.6 &9.8 &4.2\nl  
Radius$<$80" &17 &0.92 &2.36 &1.9 &0 &47 &0.2355$\pm$0.0030 &19.5 &1.6 &9.9 &7.2\nl  
Radius 80-320" &47 &0.79 &2.16 &3.6 &-5 &190 &0.2349$\pm$0.0025 &20.0 &0.3 &10.9 &5.8\nl
Radius$>$320" &18 &0.70 &1.87 &2.5 &-12 &452 &0.2349$\pm$0.0018 &20.1 &0.3 &8.4 &5.2\nl 
Near field &3 &0.89 &2.91 &1.5 &12 &283 &0.2283$\pm$0.0007 &19.0 &-0.4 &9.4 &4.4\nl
Field &19 &0.59 &1.77 &3.8 &-15 &365 &0.24--0.28 &20.2 &1.6 &4.3 &2.6\nl
Group A &6 &0.91 &2.16 &2.3 &-2 &56 &0.2366$\pm$0.0040 &20.1 &0.5 &6.6 &4.7\nl
Group B &15 &0.89 &2.44 &2.2 &2 &55 &0.2342$\pm$0.0024 &19.4 &1.4 &12.9 &8.4\nl
Group 1 &31 &0.73 &1.98 &3.9 &-12 &286 &0.2355$\pm$0.0024 &20.1 &0.1 &10.6 &5.2\nl 
Group 2 &16 &0.82 &2.28 &2.7 &-4 &224 &0.2348$\pm$0.0025 &19.7 &0.1 &10.2 &7.0\nl
\enddata
                                                                                    \tablenotetext{a}{g-R Red/Blue cut at 0.70\\~~~~~ line measures in \AA}
\tablenotetext{b}{Mean projected radius from cD in arcsec}

\end{deluxetable}

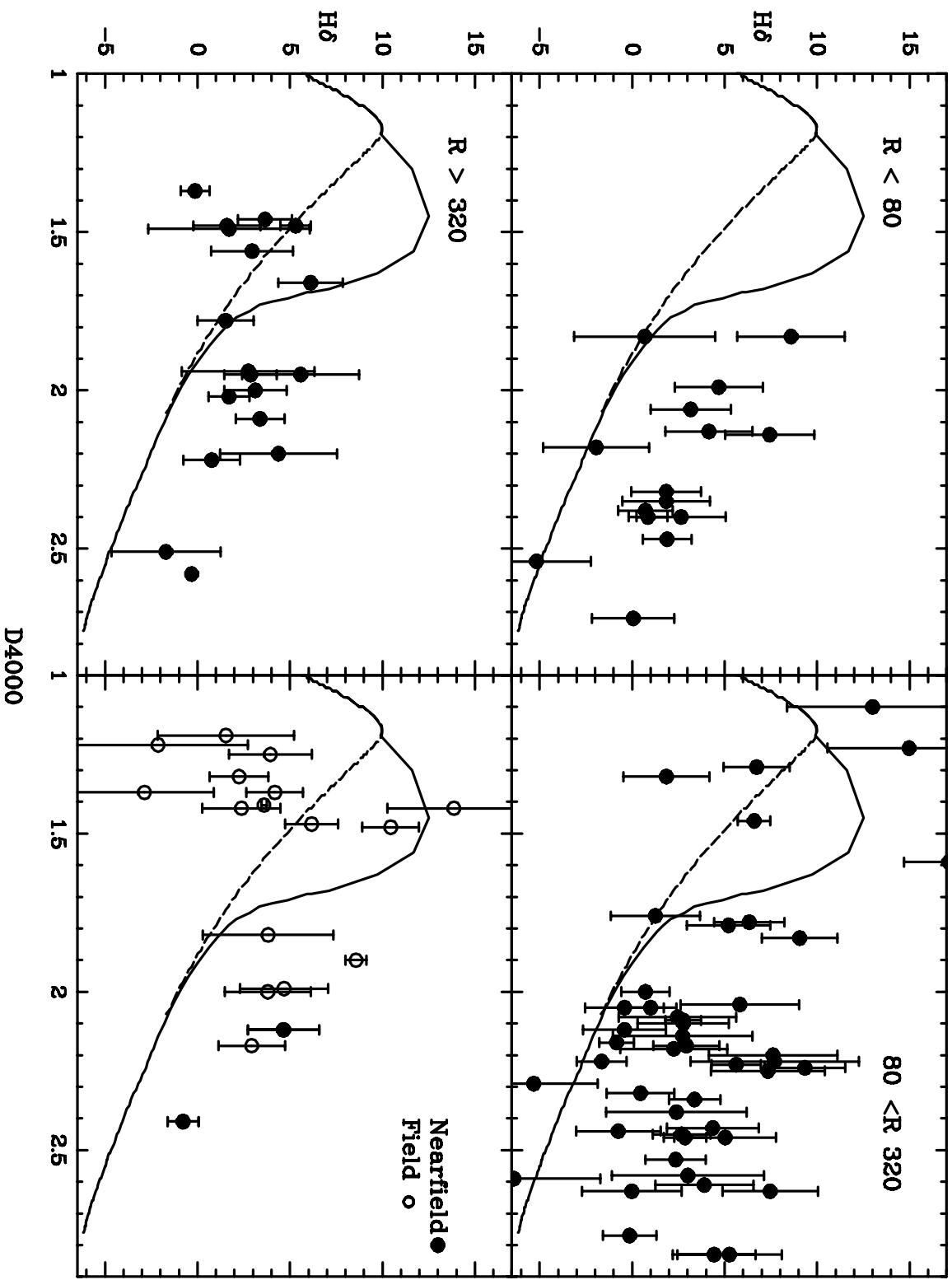
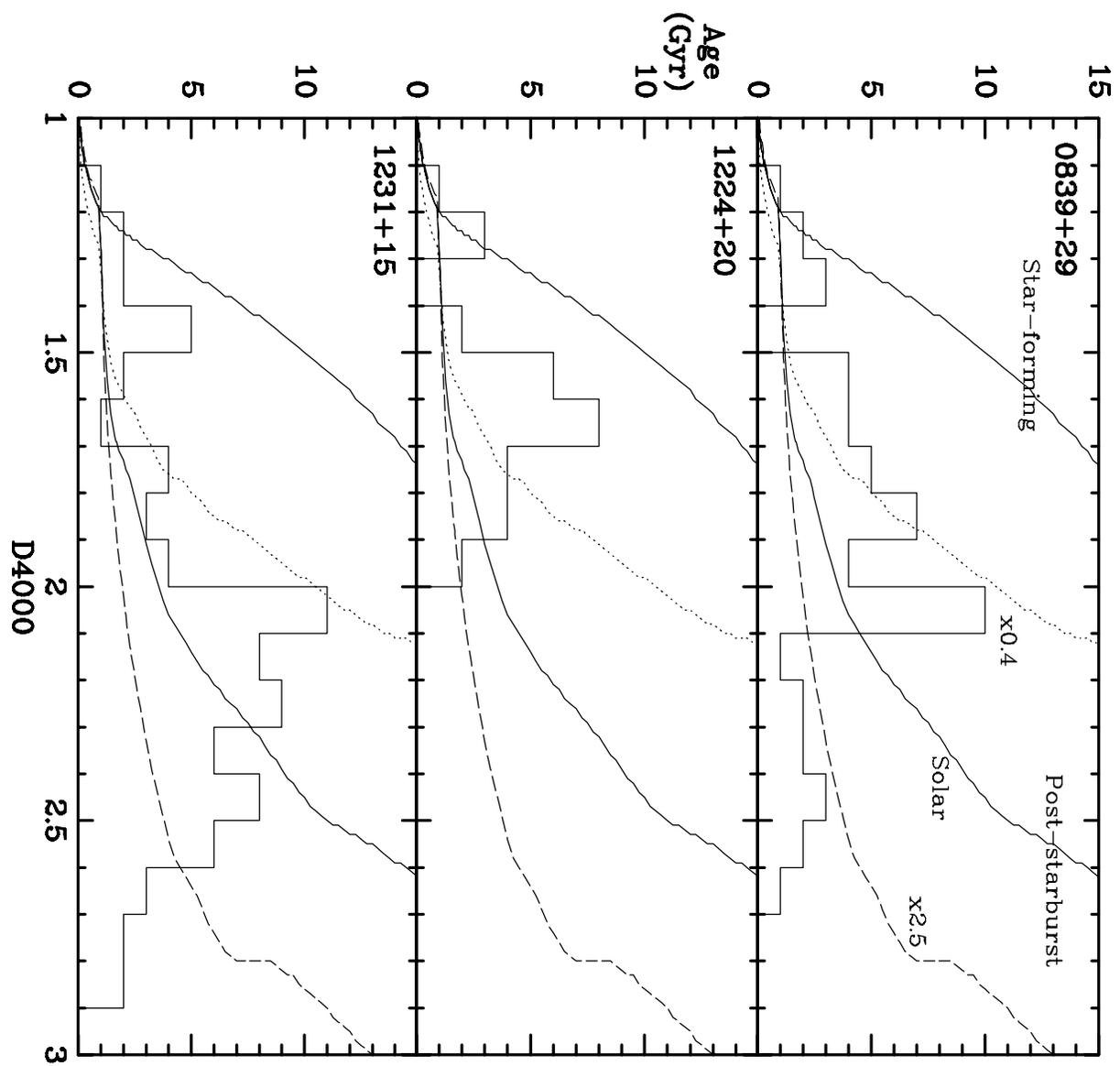
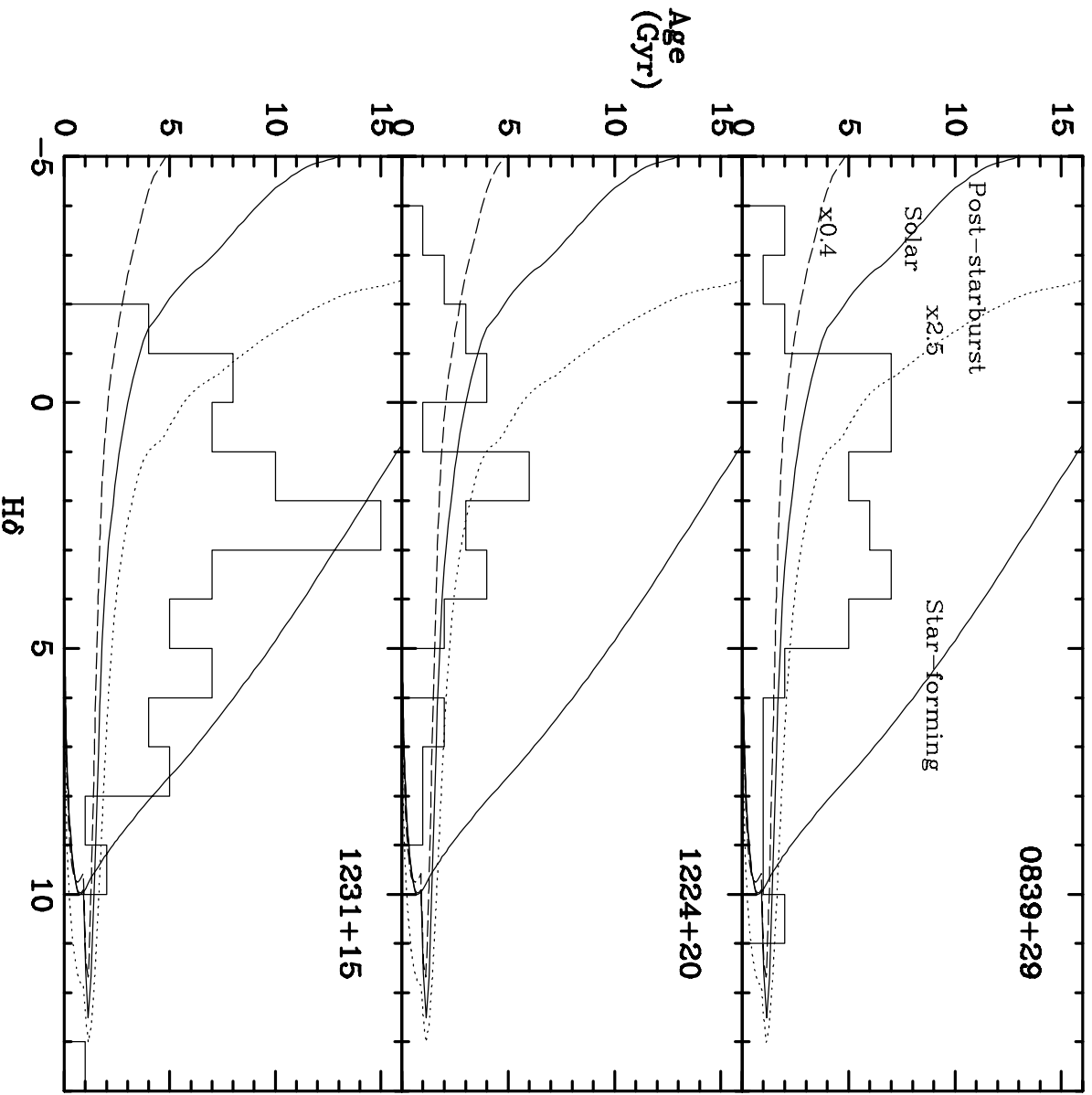
\end{document}